\documentclass[12pt]{article}
\input psfig.sty

\begin{document}

\begin{center}
{\bf CHARGE ORDERING UNDER A MAGNETIC FIELD IN THE EXTENDED
HUBBARD MODEL}
\end{center}
\begin{center}
{\it \small DUC ANH - LE$^{1,2}$, ANH TUAN - HOANG$^{1}$ and TOAN THANG - NGUYEN$^{1}$\\
\small $^{1}$Institute of Physics, P.O. Box 429 Bo Ho, \\
\small Hanoi 10 000, Vietnam\\
\small $^{2}$ Faculty of Physics, Hanoi University of Education, \\
\small 144 Xuan Thuy St, Hanoi, Viet Nam .
}
\end{center}
\begin{abstract}
 We study the charge ordering behavior under a magnetic field $H$ in the
extended Hubbard model within the coherent
potential approximation. At quarter filling, for small $H$ we
find that the relative variation of critical temperature is
quadratic with the coefficient ${\alpha}$  smaller than the one
for conventional spin-Peierls systems. For intermediate field, a
melting of the charge ordering on decreasing temperature under
fixed $H$ at various band filling is found.
\end{abstract}
{\bf 1. Introduction}\\[0.2cm]
After the discovery of a low temperature spin-gapped  phase in
$\alpha$'-NaV$_2$O$_5$$^1$, this compound has attracted a great deal of interest
 as the second example of an inorganic spin-Peierls material, even with a significantly
higher transition temperature than observed for CuGeO$_3$. The properties of
NaV$_2$O$_5$, however, have proven to be quite controversial. A number of recent
experiments do not conform  to current understanding of an ordinary spin-
Peierls system$^{2, 3}$. In particular, recent NMR measurements revealed the
appearance of two inequivalent types of V sites, V$^{4+}$ and V$^{5+}$ below the
structural transition at $T_c = 34 K$, which clearly indicates that the transition
may be driven by charge ordering (CO). In order to probe the rich electronic structure  of
NaV$_2$O$_5$ specific heat measurements and optical measurements
have been performed in magnetic field$^{3-5}$. According to the standard
theory, the magnetic field dependence of transition temperature
$T_c$  of spin-Peierls system obeys the following equation:
 $\frac{T_c(H)}{T_c(0)} = 1-\alpha [g\mu{_B}H/2k{_B}T_c(0) ]^2\,,$
where the prefactor $ \alpha$ equals to 0.44 or 0.36 depending on the
way interaction effects are taken into account $^{6,  7}$.
However, the experimental results are rather controversial as
regards the value of the prefactor $\alpha$. By polarized optical
reflectance studies one found $ 0.22 < \alpha < 0.42$ $^3$. In
contrary, specific-heat measurements in magnetic field up to 16
Tesla gave $\alpha \approx 0.092$, which is much smaller than
expected from spin-Peierls theory$^4$. Although it is not clear whether
 the charge ordering precedes or forms simultaneously with the spin-Peierls
 state, it seems certain that the physics of charge ordering must be taken
 into account, thus stimulating the research reported in this paper.
The present paper is devoted to the consideration of the effect of
the magnetic field on the CO transition temperature in the simplest
model which allows for a CO transition due to the competition
between kinetic and Coulomb energy, namely, the extended Hubbard
model (EHM) with the nearest neighbor Coulomb interaction.
To solve this problem we use coherent potential approximation (CPA),
 a simple but physically meaningful approximation which allows us to study
 the reentrant CO behavior in EHM under zero magnetic field as done in Refs. 8-9.
 Although the CPA treatment
of the Hubbard model fails in properly describing the coherent propagation of
low-energy quasiparticle in doped Mott insulators, it is not crucial in the doping
regimes explored in the present paper$^{10}$.
This
paper is organized as follows. In next section, we describe the
model and the formalism, and then in the Sec. 3 we show the
results obtained for the ($T-H$)-phase diagram. A brief
summary is given in Sec. 4.\\[0.2cm]
{\bf 2.   Model and Formalism}\\[0.2cm]
 We study the EHM in magnetic field. The Hamiltonian is given by:
\begin{equation}
H = t {\sum_{<ij>\sigma }}(c_{i\sigma }^{+}c_{j\sigma }
 + c_{j\sigma }^{+}c_{i\sigma})  +
U{\sum_{i}}n_{i\uparrow }n_{i\downarrow } +
V{\sum_{<ij>}}n_{i }n_{j}-\frac{1}{2} g\mu{_B}H{\sum_{i}}(n_{i\uparrow }-n_{i\downarrow })\,,
\end{equation}
where $c_{i \sigma} (c_{i \sigma}^+)$ annihilates (creates) an
electron with spin $\sigma$ at site $i$,  $n_{i \sigma} = c_{i
\sigma}^+c_{i \sigma}$ and $n_i = n_{i \uparrow} + n_{i
\downarrow}. <ij>$ denotes nearest neighbors, $t$ is the hopping
parameter, $U$ and $V$ are on-site and inter-site Coulomb
repulsion, respectively. The fourth term in (2.1) is the Zeeman
coupling, where $H$ is the applied magnetic field, $\mu_B$ is the
Bohr magneton, $g$ is the $g$-factor in the direction of the
magnetic field and is taken to be equal $1.98$$ ^{11}$. As we are
interested in charge ordered phase with different
occupancies on the nearest neighbor sites, we divide the cubic
lattice in two sublattices such that points on one sublattice have
only points of the other sublattice as nearest neighbors. The
sublattice is denoted by subindex $A$ or $B$: $c_{i \sigma}= a_{i
\sigma} (b_{i \sigma})$ if $i \in A \; (i \in B)$. First, we
perform a mean-field decoupling of the $V$ term in (2.1). Then, by
employing the alloy-analog approach we get a one-particle
Hamiltionian which is of the form:
\begin{eqnarray}
\tilde H=\sum\limits_{i \in A}^{} {(E_A^ +  a_{i \uparrow }^ +  a_{i \uparrow }^{}  + E_A^- a_{i \downarrow }^ +  a_{i \downarrow }^{} )}  + \sum\limits_{j \in B}^{} {(E_B^ +  b_{i \uparrow }^ +  b_{i \uparrow }^{}  + E_B^- b_{i \downarrow }^ +  b_{i \downarrow }^{} )} \nonumber \\ +
t\sum\limits_{<{ij}> \sigma }^{} {(a_{i\sigma }^ +  b_{j\sigma }  + b_{j\sigma }^ +  a_{i\sigma } )-3VNn_A n_B^{} ,}
\end{eqnarray}
\begin{equation}
E_{A/B, \sigma}^\pm = \cases{
6Vn_{B/A}\mp h \quad \textrm{with probability}\quad  1-n_{A/B,-\sigma}^\mp\,,\cr
6Vn_{B/A}\mp h+ U \quad  \textrm{with probability}\quad  n_{A/B,-\sigma}^\mp\,.\cr}
\end{equation}\\
Here $n_{A/B}^\pm$ is the averaged electron occupation number with spin up (down) in the $A/B$-sublattice, $n_{A/B}=n_{A/B}^++n_{A/B}^-$, $N$ is the number of sites in the lattice and $h=\frac{1}{2}g\mu_B H $. In CPA the averaged local Green functions for $A/B$-sublattice $\bar{G}_{A/B}$ then take the form \\
\begin{equation}
\bar{G}_{A/B}^\pm (\omega) =  \frac{2}{W^2} \left\{\omega -\Sigma_{B/A}^\pm(\omega)
- \left[ (\omega-\Sigma_{B/A}^\pm (\omega))^2 -\frac{\omega-\Sigma_{B/A}^\pm (\omega)}
{\omega-\Sigma_{A/B}^\pm (\omega)}W^2 \right]^{1/2}  \right\},\nonumber\\
\quad
\end{equation}
where we have employed the semi-elliptic density of states (DOS) for non-interacting electron, $\,\rho_0(\varepsilon) = \frac{2}
{\pi W^2}\sqrt{W^2-\varepsilon^2}$  with the bandwidth $W$.
The CPA demands that the scattering matrix vanishes on average. This yields expression for self-energy $\Sigma_{A/B}^\pm (\omega)$ of the form
 \begin{equation}
\Sigma_{A/B}^\pm(\omega) = \bar{E}_{A/B}^\pm-(6Vn_{B/A}-\Sigma_{A/B}^\pm(\omega)) \bar{G}_{A/B}^\pm
(\omega)(6Vn_{B/A} + U-\Sigma_{A/B}^\pm(\omega))\,,
\end{equation}
where   $\bar{E}_{A/B}^\pm = 6Vn_{B/A} \mp h+Un_{A/B}^\mp$.
From Eqs. (2.4)-(2.5), it is easy to obtain a system of equations for $\bar{G}_{A/B}^+$ and $\bar{G}_{A/B}^-$. For arbitrary value of electron density $n$ we denote $n_{A/B}=n\pm x,$ $n_{A}^\pm=(n_A\pm m_A)/2,$ $n_{B}^\pm=(n_B\pm m_B)/2,$ where $m_{A/B}$ is the magnetization in $A/B$-sublattice, then at temperature $T$ we have the following self-consistent system of equations for order parameters $x, m_A, m_B$ and the chemical $\mu$ for fixed $U,$ $V,$ $T,$ $h$ and $n$.
\begin{eqnarray}
n+x&=&-\frac{1}{\pi}\int_{-\infty}^{+ \infty}
d \omega f(\omega)\Im (\bar{G}_A^+(\omega)+\bar{G}_A^-(\omega))\,,\\
n-x&=&-\frac{1}{\pi}\int_{-\infty}^{+ \infty}
d \omega f(\omega)\Im (\bar{G}_B^+(\omega)+\bar{G}_B^-(\omega))\,,\\
m_A&=&-\frac{1}{\pi}\int_{-\infty}^{+ \infty}
d \omega f(\omega)\Im (\bar{G}_A^+(\omega)-\bar{G}_A^-(\omega))\,,\\
m_B&=&-\frac{1}{\pi}\int_{-\infty}^{+ \infty}
d \omega f(\omega)\Im (\bar{G}_B^+(\omega)-\bar{G}_B^-(\omega))\,.
\end{eqnarray}
Here $f(\omega) = (1 + \exp (\omega-\mu)/k_{B}T)^{-1}$ is the
Fermi function.\\ We are now interested in the phase boundary
between homogeneous ($x=0$) and charge ordered $(x\not=0)$ phases.
In this phase boundary $m_A=m_B\equiv m$ and we make following
ansatz: $\bar{G}^\pm_A(x=0,\pm m,\omega)=\bar{G}^\pm_B(x=0,\pm
m,\omega)\equiv g(\pm m,\omega)$. We find that the conditions for
the onset of CO under a magnetic field are expressed as
\begin{eqnarray}
n+m&=&-\frac{2}{\pi}\int_{-\infty}^{+ \infty}
d \omega f(\omega^-)\Im g(m,\omega)\,,\\
n-m&=&-\frac{2}{\pi}\int_{-\infty}^{+ \infty}
d \omega f(\omega^+)\Im g(-m,\omega)\,,\\
1&=&-\frac{1}{\pi}\int_{-\infty}^{+ \infty}
d \omega [f(\omega^-)\Im g'(m,\omega)+ f(\omega^+)\Im g'(-m,\omega)],
\end{eqnarray}
where $\omega^\pm=\omega\pm h$ and $g(\pm m,\omega)$ is a solution with negative imagine part of the cubic equation in the form
\begin{equation}
g^3-8\omega g^2+[16 \omega^2-4(U^2-1)]g-[16 \omega+8 U(n-1\mp m)]=0,
\end{equation}
and $g'(\pm m,\omega)$ are given by $g'(\pm m,\omega) =
\frac{\partial \bar{G}^\pm (x,m, \omega)}{\partial x} \mid_{x=0}$.
Hereafter, the bandwidth $W$ is taken to be unity for simplicity.
Setting $h=0$ and $m=0$ in Eqs. (2.10)-(2.12) we reproduce the
CPA equations for the charge ordering in EHM under zero magnetic
field in Ref. 8. For fixed temperature $T$, on-site Coulomb repulsion $U$,
 banding filling $n$ and magnetic field $H$, we have the closed system of equations
 (2.10)-(2.13) for the critical value $V$, the chemical potential $\mu$ and
  the magnetization $m$.\\[0.2 cm]
\\{\bf 3. Numerical Results and Discussion}\\[0.2cm]
We have solved numerically the system of Eqs. (2.10)-(2.13); the results can be
summarized as follows:\\
\hspace*{0.5cm}For small $H$, magnetic field decreases critical temperature and $T_c(H)$
obeys the following equation
\begin{center}
\hspace*{3cm}$\frac{T_c(H)}{T_c(0)} = 1-\alpha [g\mu{_B}H/2k{_B}T_c(0) ]^2\,,
\hspace{3cm}(3.1)$
\end{center}
In order to compare our results with experiments, we calculate the
prefactor $\alpha$ in the equation (3.1).
Fig. 1 shows relative
variation of $T_c$ as a function of the scaled magnetic field
$g\mu{_B}H/2k{_B}T_c(0)$ for different values of $U$. The inset in Fig. 1 shows the
dependence of the value $\alpha$ on the on-site Coulomb repulsion $U$ for
$
V=0.3$ and $n=\frac{1}{2}$. From our calculations for small $U$ the coefficient
$\alpha$ decreases with increasing $U$: for $0.5\leq U\leq 1.25$ we find $0.16\leq
 \alpha\leq0.20$.\\

\begin{figure}[t]
\centerline{
\psfig{figure=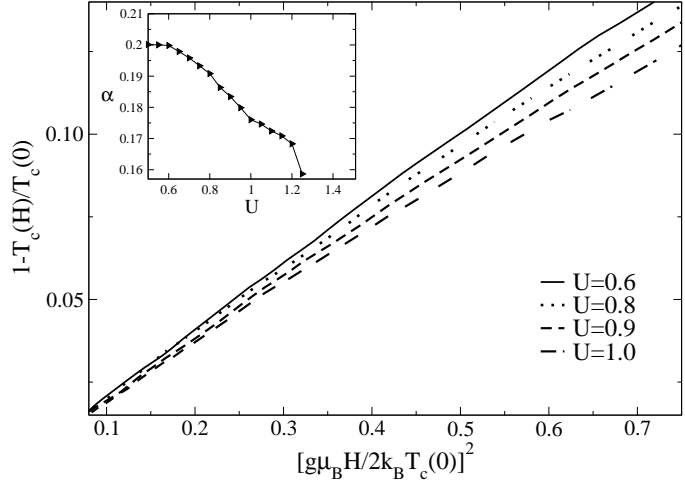,width=0.73\textwidth,angle=-90}
}
\caption{\small{ Relative variation of critical temperature as a function of the scale magnetic field
for} \small{n =0.5, V=0.3}.
\small{Inset shows the dependence of $\alpha$ on \small{U} \small{for} \small{n =0.5, V=0.3}.}}
\end{figure}
\begin{figure}[ht]
\centerline{
\psfig{figure=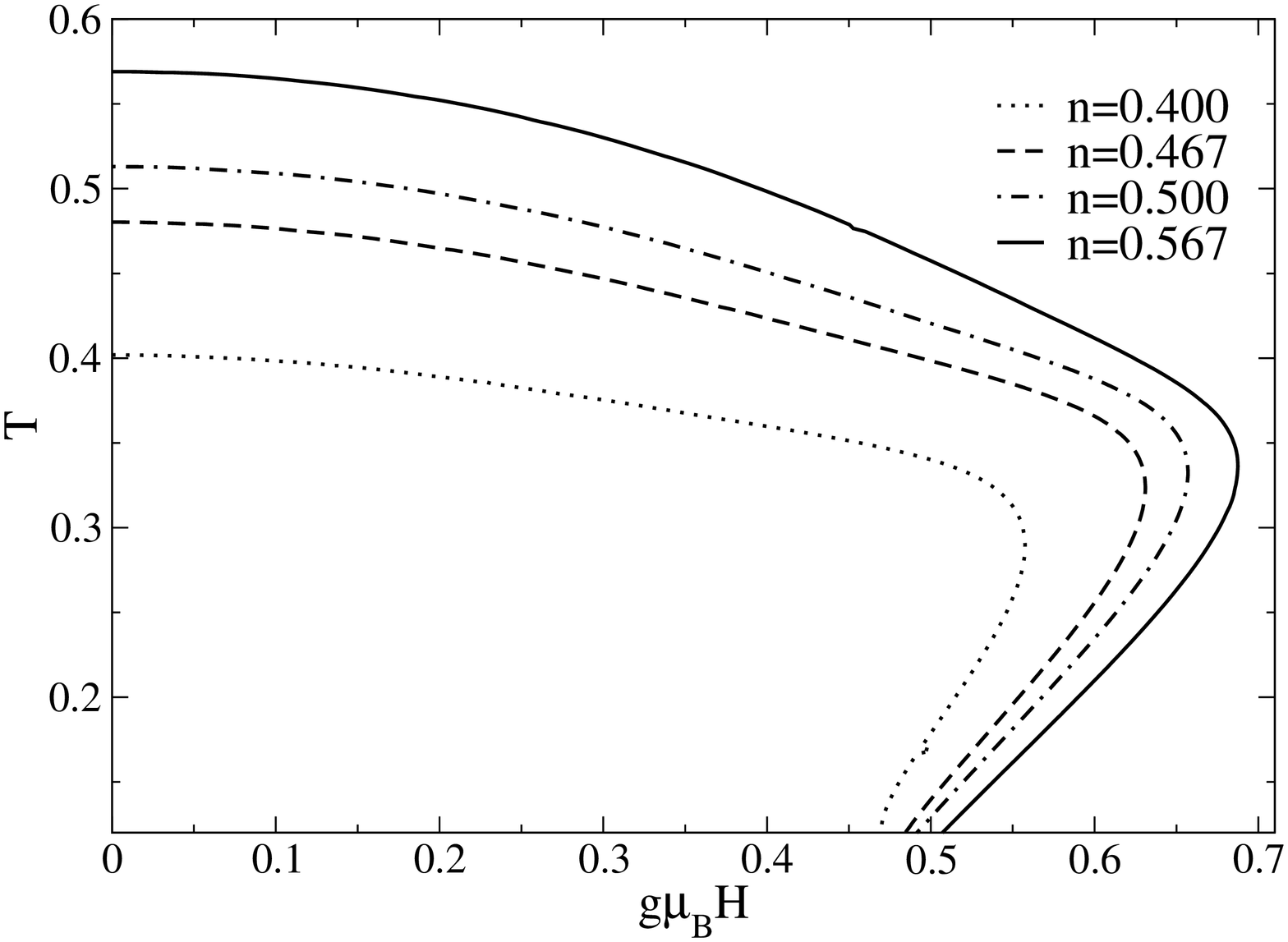,width=0.75\textwidth,angle=-90}
}
\caption{\small{(T-H)}\small{ phase diagram for} \small{U =0.5, V=0.3}
\small{at various values of \small{n}}.}
\end{figure}
As discussed in the introduction, the experimental results are controversial as
 regards the value of the coefficient $\alpha$ in NaV$_2$O$_5$, and there is
the difference between the experimental value $\alpha$ and the theoretically
  predicted one $\alpha\approx 0.36$ for spin-Peierls systems. Although this
  issue is not fully understood, it was argued by Bompardre and coworkers
in Ref. 5  that the charge density wave formation is the driving force
behind the  opening of a spin gap and the "charge" part of the transition
is mainly responsible for the $T_c(H)$ dependence, i. e. the physics of
charge ordering must be taken into account.Our calculations based on
the  EHM support this assumption. It is interesting that our results
derived from rather simple model are overall in good agreement with
experimental measurements.\\
\begin{figure}[t]
\centerline{
\psfig{figure=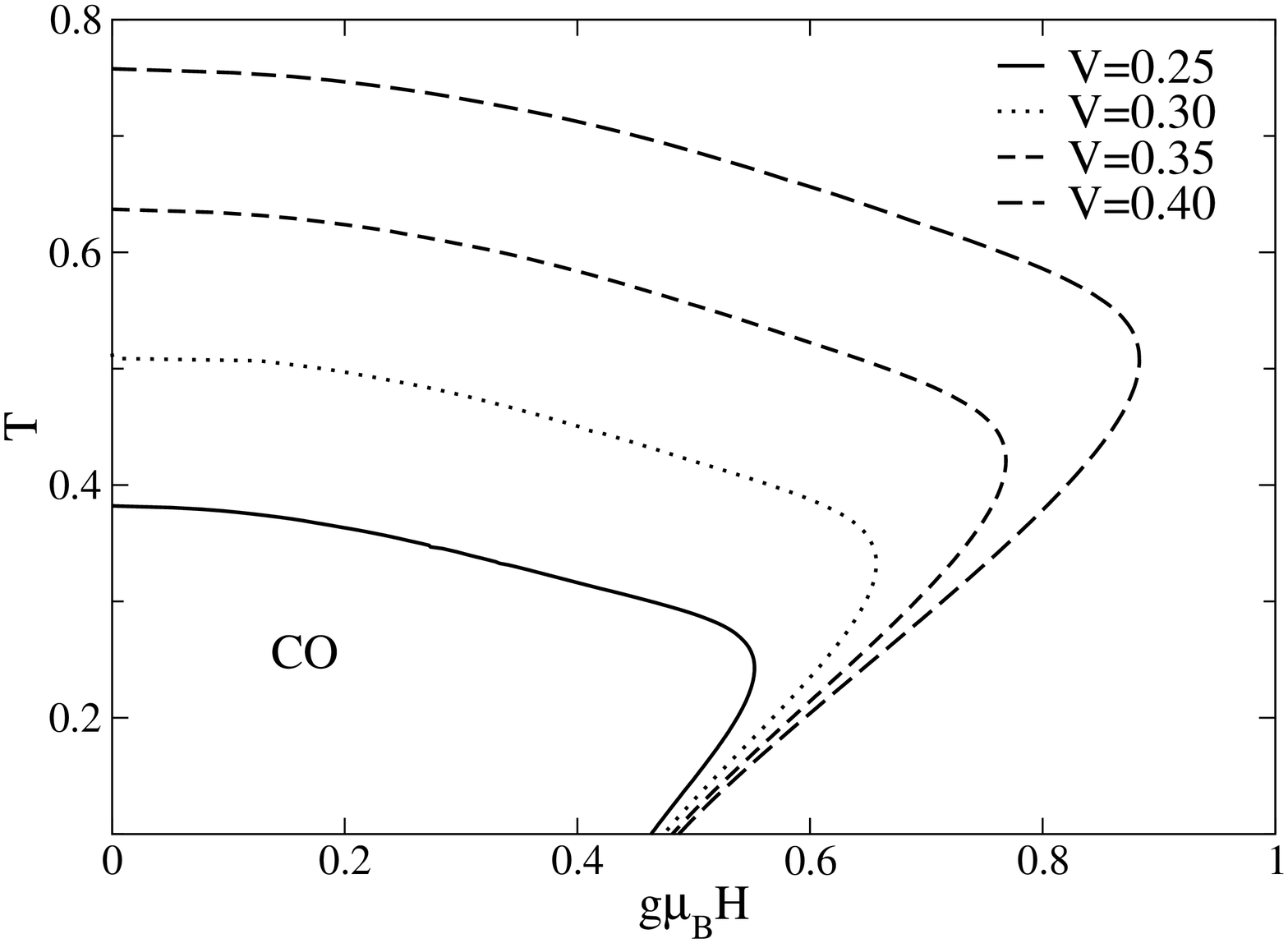,width=0.75\textwidth,angle=-90}
}
\caption{\small{(T-H)}\small{ phase diagram at quarter
filling for}
 \small{U =0.5} \small{and several values of \small{V}}.}
\end{figure}
\begin{figure*}[h]
\centerline{
\psfig{figure=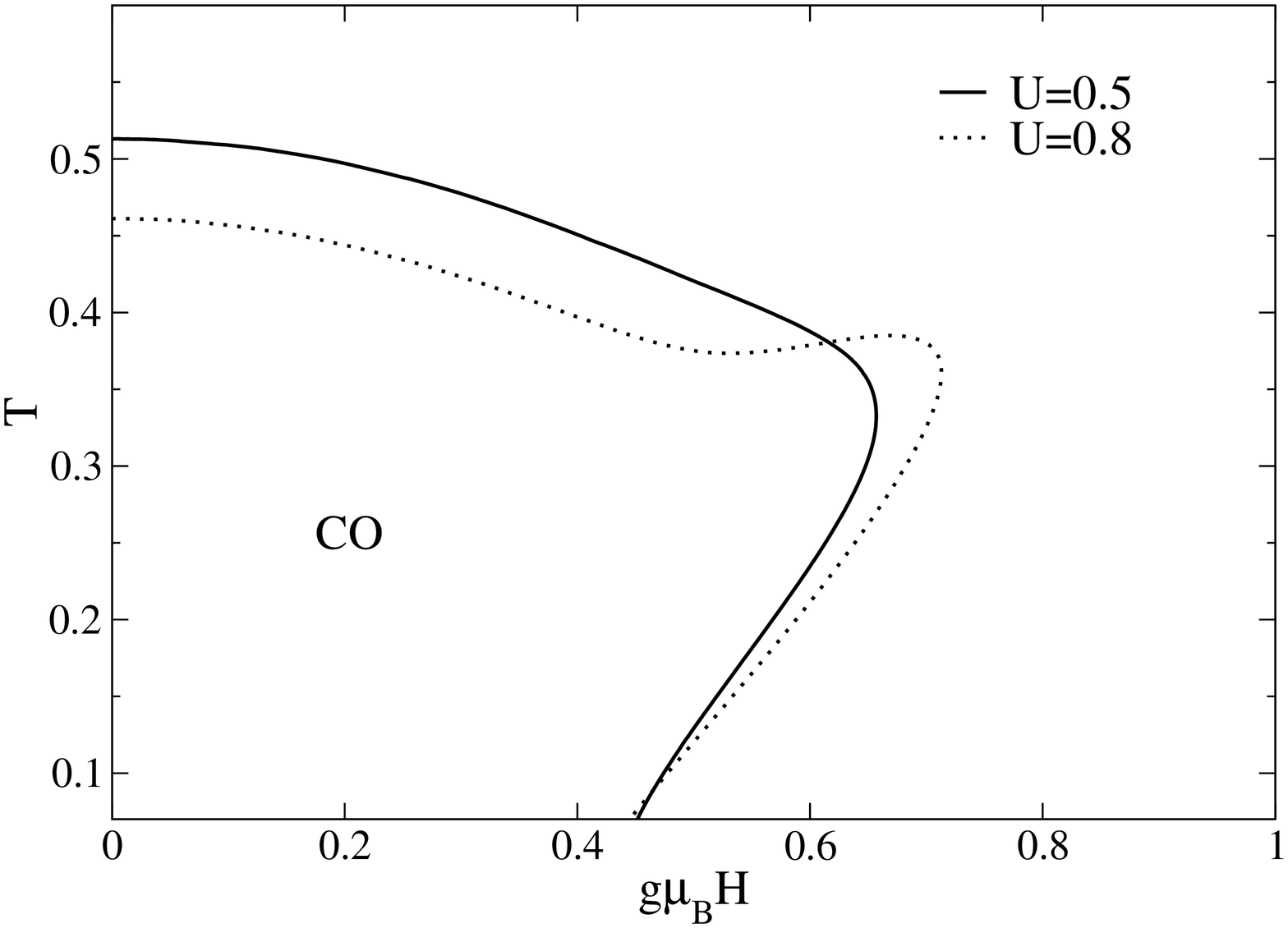,width=0.75\textwidth,angle=-90}
}
\caption{\small{(T-H)}\small{ phase diagram at quarter filling for} \small{V=0.3} \small{and different
values of \small{U}}.}
\end{figure*}
\hspace*{0.5cm}For intermediate field  the critical temperature $T_c(H)$ is not obeyed
Eq. (3.1). Furthermore, at various band filling we find that the critical $H_c$,
 as a function of temperature $T$ is found
to be non-monotous. Consequently, $\frac{dH_c}{dT}$ becomes
 positive at low temperature, i. e. reentrant CO transition with change
 of $T$ under fixed $H$ occurs, as can be clearly seen in Fig. 2 .
 In order to study the reentrant CO behavior in more detail we consider the
 ($T-H$) phase diagram for various values of the inter-site and the on-site interactions.
  The $(T-H)$ phase diagram at quarter filling for various values of the inter-site
  interaction V is displayed in Fig. 3. Reduction of the CO region with decreasing $V$
  is clearly seen. On the other hand, the reentrant charge ordered behavior is found
  for all values of $V$ in the interval $0.25<V<0.4$ with $U=0.5$.
  Fig. 4 shows the ($T-H$) phase diagram at quarter filling for several values of
   the on-site interaction $U$. The reentrant CO is clearly observed within a finite region of $h$ for $U=0.5, 0.8$.
   In our calculation the inter-site
   interaction $V$ is fixed to equal 0.3, for which the reentrant CO is not found under
    zero magnetic field $H=0$ for all above values of $U$.
    The fact that reentrant CO occurs under a magnetic field for values of $V$,
    for which reentrant CO is not observed without magnetic field is not surprised,
     since application of magnetic field causes the destruction of the charge ordering and
     at low temperature the transition field may decrease with decreasing
     temperature due to higher spin entropy of the charge ordered state$^{11, 12}$. It is
      worthy to note that recently a melting of the CO state on decreasing
      the temperature, i.e., reentrant behavior, has been found in manganites both without$^{14-17}$
      and under a magnetic field$^{12,18-20}$. On the theoretical side, to our knowledge,
         only a few studies of the reentrant CO in manganites exits$^{8,9,13,21-24}$.
    Actually we notice that except Ref. 24, most authors adopted the EHM
    with the intersite Coulomb interaction as the driving force for CO.
    Although the EHM likely lacks some important physical infredients for
    a suitable description of the manganites, the theoretical investigations,
    based on the EHM, have given a rather reasonable agreement with the experimental
    results$^{17}$ on the reentrant CO in mangannites. However, in order to quantitatively
    explain the experimental finding, it should take a more realistic model
    including the double-exchange mechanism, intersite Coulomb interaction
    and electron-phonon coupling.  \\[0.2cm]
{\bf   4. Conclusions}\\[0.2cm]
 In this paper we have applied the CPA to study the charge ordering in the
extended Hubbard model under a magnetic field. Various phase diagrams in the plane of $T$ and $H$ have shown
 and discussed. For small $H$ we find that the relative variation of critical temperature
  is quadratic with the coefficient $\alpha$ smaller than the one for conventional
  spin-Peierls systems. For small $U$ the coefficient $\alpha$ decreases with increasing
  $U$ and for $0.5\leq U \leq 1.25$ we obtained $0.16\leq \alpha\leq 0.2$.
   For intermediate field, we find a parameter region of $V$ where the model shows
   reentrant behavior in ($T-H$) phase diagram. A melting of the CO on decreasing $T$
   under fixed $H$ can be explained in terms of the higher spin entropy of charge ordered state.
The calculation presented here can also be improved by including the polaron effect. This is left for future work.\\[0.2cm]
{\bf Acknowledgments}\\[0.2cm]
 We acknowledge the referees for valuable comments, which considerably improved this work.
 This work has been supported in the part by Project 411101,
 the National Program for Basic Research on Natural Science.\\[0.2cm]
 \newpage
{\bf References}\\[0.19cm]
{\small 1. M. Isobe and Y. Veda}, {\small J. Phys. Soc. Jpn.} {\bf 65},
 {\small 1178 (1996)}.\\
{\small 2.  T. Ohama} {\small et al., Phys. Rev.} {\bf B59},
{\small 3299 (1999)}.\\
{\small 3. V. C. Long} {\small et al., Phys. Rev.} {\bf B60},
 {\small 15721 (1999)}.\\
{\small 4. W. Schnelle, Yu. Grin and R. K. Kremer},
{\small Phys. Rev. } {\bf B59}, {\small 73 (1999)}.\\
{\small 5. S. G. Bompardre } {\small et al., Phys. Rev.} {\bf B61},
 {\small R 13321 (2000)}.\\
{\small 6.  L. N. Bulaevskii, A. I. Buzdin, and D. I. Khomskii},
{\small  Solid State Commun.} {\bf 27}, {\small 5 (1978)}.\\
{\small 7. M. C. Cross}, {\small Phys. Rev.} {\bf B20}, {\small 4606 (1979)}.\\
{\small 8. Hoang Anh Tuan}, {\small Mod. Phys. Lett.} {\bf B15}, {\small 1217 (2001)}.\\
{\small 9. A. T. Hoang and P. Thalmeier},
{\small J. Phys.: Cond. Mat.} {\bf 14, }{\small 6639 (2002)}.\\
{\small 10. The authors are grateful to the referee for pointing out
 this physical issue}.\\
{\small 11. Ogawa} {\small et al., J. Phys. Soc. Jpn.} {\bf 55},
 {\small 2129 (1986)}.\\
 {\small 12. M. Tokunaga} {\small et al., Phys. Rev.} {\bf B57},
 {\small 5259 (1998)}.\\
{\small 13. R. Pietig, R. Bulla and S. Blawid},
{\small Phys. Rev. Lett.} {\bf 82}, {\small 4046 (1999)}.\\
{\small 14. T. Kimura} {\small et al., Phys. Rev. Lett.} {\bf 79},
 {\small 3720 (1997)}.\\
 {\small 15. T. Chatterji} {\small et al., Phys. Rev. } {\bf B61},
 {\small 570 (2000)}.\\
{\small 16. J. Q. Li} {\small et al., Phys. Rev. } {\bf B64},
 {\small 174413 (2001)}.\\
{\small 17. J. Dho},
{\small et al., J. Phys. Cond. Mat.} {\bf 13, }{\small 3655 (2001)}.\\
{\small 18. Y. Tomioka} {\small et al., Phys. Rev. } {\bf B53},
 {\small R1689 (1996)}.\\
 {\small 19. M. Prellier} {\small et al., Phys. Rev. } {\bf B62},
 {\small R16337 (2000)}.\\
 {\small 20. E. R. Buzin} {\small et al., Appl. Phys. Lett.} {\bf 79},
 {\small 647 (2001)}.\\
{\small 21. Q. Yuan and P. Thalmeier,} {\small Phys. Rev. Lett.} {\bf 83},
 {\small 3502 (1999)}.\\
{\small 22. C. S. Hellberg,} {\small J. Appl. Phys. } {\bf 89},
 {\small 6627 (2001)}.\\
{\small 23. D. Khomskii,} {\small Physica} {\bf B280},
 {\small 325 (2000)}.\\
 {\small 24. Q. Yuan and T. Kopp,} {\small Phys. Rev.} {\bf B65},
{\small 174423 (2002)}.
\end{document}